\documentclass[onecolumn,showpacs,preprintnumbers,elsart]{revtex4}
\usepackage{makeidx}
\usepackage{amssymb}
\usepackage{amsmath}
\usepackage{mathrsfs}
\usepackage{graphicx}
\usepackage{dcolumn}
\usepackage{bm}
\usepackage[center]{subfigure}
\usepackage{color}

\begin{document}

\title{Cross-symmetry breaking of two-component discrete dipolar matter-wave solitons }
\author{Yongyao Li$^{1}$, Zhiwei Fan$^{2}$, Zhihuan Luo$^{2}$, Yan Liu$^{2}$, Hexiang He$^{1}$, Jiantao L\"{u}$^{1}$, Jianing Xie$^{1}$, Chunqing Huang$^{1}$}
\email{Chunqinghuang@qq.com}
\author{Haishu Tan$^{1}$}
\affiliation{$^{1}$School of Physics and Optoelectronic Engineering, Foshan University, Foshan 528000, China\\
$^{2}$Department of Applied Physics, South China Agricultural University,
Guangzhou 510642, China }

\begin{abstract}
We study the spontaneous symmetry breaking of dipolar Bose--Einstein
condensates trapped in stacks of two-well systems, which may be effectively built as
one-dimensional trapping lattices sliced by a repelling
laser sheet. If the potential wells are sufficiently deep, the system is modeled
by coupled discrete Gross--Pitaevskii equations with nonlocal self- and
cross-interaction terms representing dipole--dipole interactions. When the dipoles are not polarized
perpendicular or parallel to the lattice, the cross-interaction is asymmetric, replacing the familiar symmetric two-component solitons with a new species of cross-symmetric or -asymmetric ones. The
orientation of the dipole moments and the interwell hopping rate strongly
affect the shapes of the discrete two-component solitons as well as the characteristics of the cross-symmetry breaking and the associated phase transition. The sub- and super-critical types of cross-symmetry breaking can be controlled by either the hopping rate between the components or the total norm of the solitons. The effect of the interplay between the contact nonlinearity and the dipole angle on the cross-symmetry breaking is also discussed.\\
\textbf{Keywords} Discrete matter-wave solitons, two-component systems, dipole--dipole interactions, cross-symmetry breaking.
\end{abstract}

\pacs{42.65.Tg; 03.75.Lm; 05.45.Yv}
\maketitle

\section{Introduction}

%\begin{figure}[tbp]
%\centering{\label{fig1a} \includegraphics[scale=0.6]{1a.eps}}
%\caption{(Color online) A sketch of the stack of quasi-1D dipolar
%condensates.}
%\label{sketch}
%\end{figure}
Dipolar Bose--Einstein condensates (BECs) composed of polar atoms and molecules
are the subject of active and broad research in atomic and low-temperature physics. This type of
BEC, which is dominated by anisotropic long-range magnetic or electric dipole--dipole
interactions (DDIs), differs significantly from typical BECs, whose
intrinsic dynamics is driven by isotropic contact interactions induced by
$s$-wave scattering. Studies of dipolar BECs have produced many fascinating experimental and theoretical results, which have been summarized in reviews and references therein \cite{Griesmaier2007,Lahaye20092,Baranov3}.

One important advantage of the dipolar BEC is its excellent tunability. In addition to the fact that the atomic or molecular dipoles can be polarized by external dc electric or magnetic fields, the sign of the DDI
can be switched by a rotating ac field \cite{Giovanazzi2002}. Moreover, atoms or molecules carrying field-induced magnetic or electric moments placed in spatially nonuniform fields are particularly interesting \cite{YLi2016}.
These tunable properties indicate the vast potential offered by dipolar BECs for
fundamental and applied studies. One significant direction in these studies
is the simulation of various phenomena that occur in a more complex form in
other physical media, such as ferrofluidity \cite{Lahaye2009,Saito2009,Kadau2016,Richter2005}, rotons \cite{rotons,klawunn2008,Wilson2008,Hufngl2011,Zhao2015}, Faraday waves \cite{Nath2010,Lakomy2012}, supersolids \cite{Buhler2011,Maluckov2012}, anisotropic superfluidity \cite{Ticknor2011} and anisotropic
collapse \cite{collapse2002,Bortolotti,Ticknor,Lahaye2008}, mesoscopic drops stabilized by quantum fluctuations \cite{drops,KTXI2016,Wachtler2016,Baillie2016}, and others
\cite{Klawunn2009,Muller2011,Wilson2011,Gawryluk2011,Zhao2015}.

Another remarkable ramification is the use of collective nonlinear modes in
dipolar BECs to create matter-wave solitons. This topic was
originally introduced in nonlinear optics with nonlocal media \cite{
nonlocal-reviews,Peccianti2012,Krolikowski2000,Skupin2006}. A noteworthy finding is that DDIs can help stabilize multidimensional matter-wave solitons \cite{2Dsoliton,Tikhonenkov2008}. Various forms of bright
\cite{bright,Abdullaev2013,Raghunandan2015,Adhikari2014,Huang2015,Guihua2017}, dark \cite{dark,Bland2015,Pawowski2016}, vortex \cite{vortex,Tikhonenkov22008}, and discrete \cite{discrete,Gligoric2010,Huaiyu2016,discrete2} solitons were predicted in dipolar condensates. Very recently, stable two-dimensional (2D) solitons were predicted in a dipolar BEC with spin--orbit coupling \cite{SOC,Xunda2016,Yongyao2017}. It was demonstrated that the DDI can create solitons not only in BECs but in an
ultracold bosonic gas of the Tonks--Girardeau type \cite{TG}.

Very recently, dipolar matter-wave solitons were studied in a two-component discrete system \cite{Zhiwei2017}. A sketch of this system is displayed in Fig. \ref{sketch}(a). It was found that when the dipoles are not polarized perpendicular or parallel to the lattice, the cross-interaction, which is induced by the DDIs between two lines, is asymmetric [in Fig. \ref{sketch}(b), the left and right sides of the $n$-th lattice on the upper line feel attractive and repulsive DDIs, respectively, from the lower line]. This asymmetric cross-interaction gives rise to an asymmetric nonlocal cross-phase-modulation (XPM) term in the discrete Gross--Pitaevskii equation (GPE) and brings about a new type of symmetry, which is called cross-symmetry, in two-component discrete solitons [see Eq. (\ref{newrelationship}) below]. Typically, when the dipole angle $\theta$ is near $35.3^{\circ}$ [see Eq. (\ref{theta1}) below], two types of cross-symmetry, on-site (the cross-symmetric axis is at a lattice site) and off-site (the cross-symmetry axis is at the midpoint between two sites) cross-symmetry, are induced by the asymmetric cross-interaction of the DDI. However, in a broad range of dipole angles, cross-symmetry breaking, which replaces the off-site cross-symmetry, was not discussed in this paper. The process of generating cross-asymmetry solitons from cross-symmetry solitons remains unclear.

The objective of the present work is to investigate spontaneous cross-symmetry breaking and the associated phase transition, also known as the symmetry-breaking bifurcation, of discrete solitons supported by the asymmetric cross-interaction (i.e., nonlocal asymmetric XPM) in two-component systems. Spontaneous symmetry breaking is generally a ubiquitous phenomenon that occurs in broad areas of nonlinear physics \cite{SSB}. As solitons are an important nonlinear phenomenon, a natural subject of analysis is spontaneous symmetry breaking of solitons in symmetric systems. In particular, many theoretical and experimental results on this subject have been reported in optical and matter-wave nonlinear settings, where the symmetry is frequently provided by dual-core or double-well structures \cite{SSB2}. Symmetry breaking of solitons in systems with local nonlinearity has been studied in detail theoretically \cite{Wright1989,Pare1990,Akhmediev1993,Chu1993,Crespo1993,Matuszewski2007,Tsofe2007,Adhikari2009,Sakaguchi2011,Li2011,Merhasin2005}; the systems include discrete systems, which are represented by parallel arrays of coupled waveguides \cite{Herring2007,Yliu2017}.

Spontaneous symmetry-breaking of solitons in systems with nonlocal nonlinear media is also an interesting problem, as the nonlocality strongly affects the outcome of spontaneous symmetry breaking when the nonlinearity strength exceeds a critical value \cite{SSB2}. To date, only few works have addressed this topic. In particular, the symmetry-breaking transformation of optical solitons in a dual-core coupler with nonlocal thermal nonlinearity was considered in Ref. \cite{Shi2012}. Unlike the optical systems, the dipolar BEC dual-core setting exhibits not only intracore nonlocal nonlinearity but also an intercore DDI, which makes the situation essentially different. The differences were first shown in a model for an effectively one-dimensional (1D) dual-core coupler filled with a dipolar condensate \cite{YLi2013}. In that work, sub- and supercritical symmetry breaking, (i.e., symmetry-breaking phase transitions of the first and second kinds, respectively) could be induced by competition between the inter- and intracore DDIs, respectively. However, the analysis in Ref. \cite{YLi2013} considered a single polarization of the dipoles, namely, along the cores. In fact, an external magnetic field can polarize the dipoles in any direction, which offers an additional tool for working with dipolar systems. The aim of the present work is to explore the use of this degree of freedom to control the shape of discrete two-component solitons and the characteristics of the phase transition (symmetry-breaking bifurcation) in them. In particular, one of its essential feature is that, unless the dipole moments are oriented
strictly perpendicular or parallel to the system's axis, the shapes of the solitons become irregular, i.e., spatially asymmetric, necessitating a modification of the definition of symmetry (and asymmetry) between the soliton's components, replacing it with cross-symmetry [see Eq. (\ref{newrelationship}) below]. Moreover, spontaneous cross-symmetry breaking in discrete solitons supported by asymmetric cross-interactions has not been studied before.

The rest of the paper is structured as follows. The model is introduced in Sec.
II, and the cross-symmetry breaking in discrete two-component solitons controlled by the
orientation of the dipoles is studied in Sec. III. The paper is concluded in Sec. IV.

\section{The model}

\begin{figure}[tbp]
\centering{\label{fig1a} \includegraphics[scale=0.5]{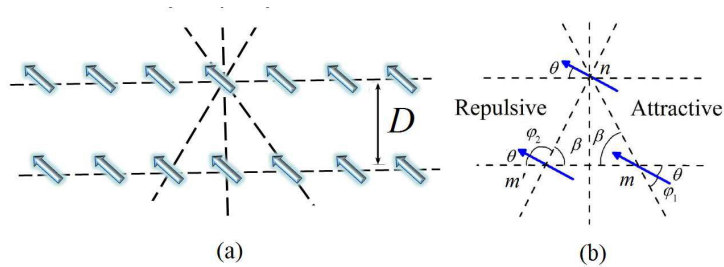}}
\caption{(Color online) (a) Dipoles in two-component discrete lattices are oriented by an external field in an arbitrary direction $\protect\theta $ with respect to the horizontal (dashed) line. $D$ is the distance between the two dashed lines. (b) Typical example of the contribution of the asymmetric nonlocal cross-interaction. In this sketch, if $\varphi_{1}<54.7^{\circ}$ and $\varphi_{2}>54.7^{\circ}$, the $m\leftrightarrow n$ and $m'\leftrightarrow n$ DDIs are attractive and repulsive, respectively, and produce the asymmetric nonlocal cross-interaction in this two-component system. }
\label{sketch}
\end{figure}

We consider a chain of two-well systems into which a dipolar BEC is
loaded, as schematically shown in Fig. \ref{sketch}(a). It can be built as
the usual quasi-1D lattice \cite{Markus,Morsch2006}, cut into a pair
of parallel scalabilities by an additional repulsive (blue-detuned) laser
sheet. Configurations with different orientations of the dipoles
with respect to the lattice are also shown in Fig. \ref{sketch}(a).

In the tight-binding approximation \cite{tight,Alfimov2002,Panos} and using Wannier-like functions \cite{discrete,discrete2,Abdullaev2001}, the mean-field
dynamics of the condensate in this system is governed by a two-component
dimensionless discrete GPE \cite{Zhiwei2017}:
\begin{eqnarray}
&&i{\frac{d}{dt}}\tilde{\psi}_{n}=-{\frac{C}{2}}(\tilde{\psi}_{n+1}+\tilde{%
\psi}_{n-1})+\left[ \sigma |\tilde{\psi}_{n}|^{2}+\sum_{m\neq n}\left(
F_{nm}|\tilde{\psi}_{m}|^{2}+G_{nm}|\tilde{\phi}_{m}|^{2}\right) \right]
\tilde{\psi}_{n}-J\tilde{\phi}_{n}  \notag \\
&&i{\frac{d}{dt}}\tilde{\phi}_{n}=-{\frac{C}{2}}(\tilde{\phi}_{n+1}+\tilde{%
\phi}_{n-1})+\left[ \sigma |\tilde{\phi}_{n}|^{2}+\sum_{m\neq n}\left(
F_{nm}|\tilde{\phi}_{m}|^{2}+G_{mn}|\tilde{\psi}_{m}|^{2}\right) \right]
\tilde{\phi}_{n}-J\tilde{\psi}_{n}.  \label{baseEq}
\end{eqnarray}%
Here, $C$ and $J$ are the coupling constants
(determined by the respective hopping rates) along the lattice and between
the wells, respectively; $\sigma $ is the strength of the contact nonlinearity, and $%
F_{nm}$ and $G_{nm}$ are DDI\ kernels that account for the
nonlocal self- and cross-interactions, respectively, in the coupled GPEs:
\begin{eqnarray}
&&F_{nm}=%
\begin{cases}
0, & (m=n) \\
{(1-3\cos ^{2}\theta )/|m-n|^{3}} & (m\neq n)%
\end{cases}%
,  \label{F} \\
&&G_{nm}=%
\begin{cases}
(1-3\sin ^{2}\theta )/D^{3} & (m=n) \\
{\left[ 1-3\cos ^{2}\varphi _{1}\right] /[D^{2}+(m-n)^{2}]^{3/2}} & (m<n) \\
{\left[ 1-3\cos ^{2}\varphi _{2}\right] /[D^{2}+(m-n)^{2}]^{3/2}} & (m>n)%
\end{cases}%
,  \label{G}
\end{eqnarray}%
where $D$ is the scaled interwell distance (vertical onsite distance), $\theta $ is the angle between
the dipole orientation and horizontal axis [see Fig. \ref{sketch}(a)], $%
\varphi _{1}=\beta -\theta $, and $\varphi _{2}=\pi -(\beta +\theta )$ [see
Fig. \ref{sketch}(c)], with $\beta \equiv \arccos \left( |m-n|/\sqrt{%
D^{2}+(m-n)^{2}}\right) $.

Typically, when $\theta=0$ or $\pi/2$, $G_{nm}$ is a symmetric matrix that satisfies $G_{nm}=G_{mn}$. For $0<\theta<\pi/2$, $G_{nm}$ is an asymmetric matrix satisfying $G_{nm}\neq G_{mn}$. This asymmetry is contributed by an asymmetric nonlocal cross-interaction [Fig. \ref{sketch}(b)]. In Ref. \cite{Zhiwei2017}, we find that when $D<0.7$, an off-site cross-symmetric soliton can be produced inside a narrow window near $\theta\approx0.196\pi=35.3^{\circ}$ [see Eq. (\ref{theta1}) below]. In this work, we consider the cross-symmetry breaking induced by the asymmetric $G_{nm}$; for convenience, we assume that the interwell distance $D$ is equal to the horizontal distance between two adjacent lattice sites, i.e., $D=1$, throughout the paper. In the following section, we will study the cross-symmetry breaking of a fundamental soliton from $\theta=0$ to $\pi/2$ by numerical simulations. Recently, a configuration similar to Eq. (\ref{baseEq}) was considered as an Ising model with long-range interactions, which does not include horizontal hopping, i.e., with $C=0$ \cite{Ising}.

Stationary states are searched for in the usual form,
\begin{equation}
(\tilde{\psi}_{n},\tilde{\phi}_{n})=(\psi _{n},\phi _{n})e^{-i\mu t},
\label{stationary}
\end{equation}%
where $(\psi _{n},\phi _{n})$ are stationary wave functions, and $\mu $ is a
real chemical potential. Two-component solitons are characterized by their
total norm,%
\begin{equation}
P=P_{\psi }+P_{\phi }\equiv \sum_{n=-N/2}^{n=N/2}\left( |\tilde{\psi}%
_{n}|^{2}+|\tilde{\phi}_{n}|^{2}\right) ,  \label{P}
\end{equation}%
which is a dynamical invariant of Eq. (\ref{baseEq}). According to Ref. \cite{discrete}, if we assume that the width of the transverse confinement is $\sim5$ $\mu$m, and the period of the optical lattice can be readily set to 5 $\mu$m, we conclude that $P=1$ may correspond to $\sim1000$ atoms in the condensates.

As mentioned above, for $\theta =0$ or $\pi /2$, the matrix $G_{nm}$ given by Eq. (\ref{G}) is
symmetric. Obviously, at $\theta =0$, the \textit{vertical }(also called interwell) interaction,
i.e., the onsite DDI between condensate droplets trapped in the two wells,
is repulsive, whereas the \textit{horizontal} DDI within each sublattice is
attractive. At $\theta=\pi/2$, the situation is the opposite; the onsite DDI in the two wells is attractive, whereas the horizontal DDI is repulsive.

When $0<\theta <\pi /2$, the symmetry of the matrix is broken, which
may contribute to an asymmetric nonlocal cross-interaction [Fig. \ref%
{sketch}(b)]. In the interval $\theta \in (0,\pi /2)$, there are two
well-known special angles:
\begin{equation}
\theta _{1}=\arcsin \left( 1/\sqrt{3}\right) \approx 0.196\pi \approx
35.3^{\circ },  \label{theta1}
\end{equation}%
\begin{equation}
\theta _{2}=\arccos \left( 1/\sqrt{3}\right) \approx 0.304\pi \approx
54.7^{\circ }.  \label{theta2}
\end{equation}%
When $\theta =\theta _{1}$, the vertical DDI vanishes, and the horizontal DDI
remains attractive; when $\theta =\theta _{2}$, the horizontal DDI vanishes, and the vertical DDI is attractive. Even though the DDI between the two parallel lattices is anisotropic, its total effect remains attractive. Therefore, symmetry breaking may occur in the ground-state solution of the system. At $\theta =0$ or $\pi /2$, the shapes of the ground-state solitons (which include symmetric and asymmetric ones with respect to the two components) that are produced by the system are spatially even, as mentioned above and shown below in Figs. \ref{SolitonExp}(a1), (a2), and (d); i.e., they obey the spatial symmetry condition,
\begin{equation}
\phi _{-n}=\phi _{n},\psi _{-n}=\psi _{n}.  \label{relationship}
\end{equation}%
However, when $\theta\neq0$ or $\pi /2$, $G_{nm}$ is not
a symmetric matrix [see Eq. (\ref{G})]; hence, the shapes of the two
components are not spatially even. In particular, it is easy to see that,
in this case, the symmetry condition with respect to the two
components actually takes the form of the \textit{cross-symmetry}, as
clearly shown in Figs. \ref{SolitonExp} (b1) and (c):
\begin{equation}
\phi _{-n}=\psi _{n},  \label{newrelationship}
\end{equation}%
which is different from Eq. (\ref{relationship}). To quantify the symmetry breaking of solitons in the system, we define the usual measure of the asymmetry between the two components in terms of their norms [cf. Eq.
(\ref{P})]:
\begin{equation}
\eta ={\frac{|P_{\psi }-P_{\phi }|}{P_{\psi }+P_{\phi }}}.  \label{eta}
\end{equation}%
In the following section, we study discrete solitons in this system and the
cross-symmetry breaking between their two components (i.e., between the parallel sublattices)
in the range $0\leq \theta \leq \pi /2$ using numerical methods.

\section{Spontaneous cross-symmetry breaking of two-component discrete
solitons}

\subsection{System with $\protect\sigma =0$ (no contact interactions)}

\begin{figure}[tbp]
\centering{\label{fig2a} \includegraphics[scale=0.35]{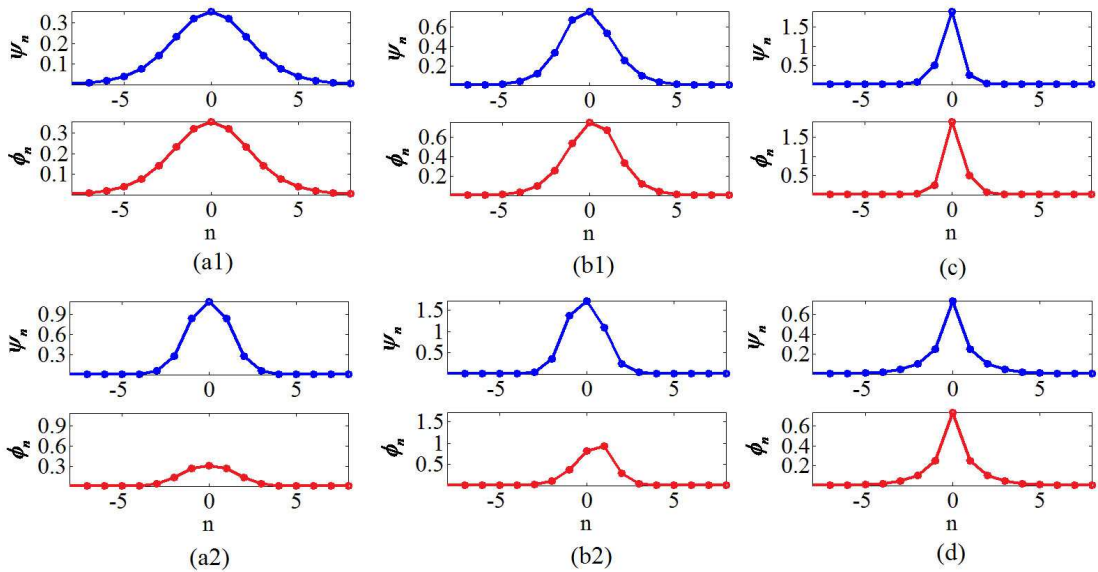}}
\caption{(Color online) Typical examples of discrete two-component solitons
obtained by the imaginary-time method. (a1,a2) Symmetric and
asymmetric solitons (with respect to the two components) for $(P,J,\protect%
\theta )=(1,1,0)$ and $(3,1,0)$, respectively. (b1,b2) Symmetric and
asymmetric solitons for $(P,J,\protect\theta )=(3,1,\arcsin (1/\protect\sqrt{%
3}))$ and $(8,1,\arcsin (1/\protect\sqrt{3}))$, respectively. (c,d)
Symmetric solitons for $(P,J,\protect\theta )=(8,1,\arccos (1/\protect\sqrt{3%
}))$ and $(P,J,\protect\theta )=(1.4,1,\protect\pi /2)$, respectively. Note
that the spatial shape of the solitons is not even except when $%
\protect\theta =0$ in (a1) and (a2); see the text for details.}
\label{SolitonExp}
\end{figure}

Ground-state discrete solitons, which always feature a site-centered \textit{%
unstaggered} (smooth) shape \cite{Panos}] and may be both symmetric and
asymmetric with respect to the two components, are obtained from Eq. (\ref%
{baseEq}) by the imaginary-time method (ITM)\cite%
{Chiofalo,Jianke1,Jianke2}. We arrange two types of initial guess for the ITM, symmetric and asymmetric, which refer to equal and unequal norm distributions, respectively. Before the symmetry-breaking bifurcation, both the symmetric and asymmetric initial guesses can generally produce only the symmetric state; however, after the symmetry-breaking bifurcation, the symmetric and asymmetric guesses produce symmetric and asymmetric states, respectively. The stabilities of the output symmetric or asymmetric states are verified by direct simulations of Eq. (\ref{baseEq}) realized by the four-step Runge--Kutta method. The stable state can propagate stably for a sufficiently long time. There are two types of symmetry-breaking bifurcation: sub- and supercritical. In the subcritical symmetry-breaking bifurcation (which is tantamount to the phase transition of the first kind), the system exhibits branches of asymmetric states that emerge as unstable ones, going at first backward from the bifurcation point and becoming stable after turning forward. Therefore, there is a small overlap (bistable) area of stable symmetry and asymmetry states; in this area, both the symmetric and asymmetric guesses can produce a stable output solution via the ITM. In the supercritical symmetry-breaking bifurcation (which is tantamount to the phase transition of the second kind), the asymmetric branches emerge as stable ones and immediately go in the forward direction.

To focus on the cross-symmetry breaking induced solely by the DDIs, here
we first remove the contact nonlinearity; i.e., we set $\sigma =0$ in Eq. (\ref{baseEq}) in this section.
Therefore, the remaining control parameters are the total norm $P$ [see Eq. (%
\ref{P})], interwell hopping rate $J$ in Eq. (\ref{baseEq}), and
orientation angle $\theta $. The horizontal hopping rate $C$ is fixed as 1 throughout the paper.

Figure \ref{SolitonExp} shows typical examples of discrete solitons found for the dipole polarization
angles $\theta =0$, $0.196\pi $, $0.304\pi $, and $\pi /2$. For $\theta =0$ and $0.196\pi $, both stable
symmetric and asymmetric solitons were found for smaller and larger
values of $P$, respectively [Figs. \ref{SolitonExp}(a1), (a2), (b1), and (b2)]. For $%
\theta =0.304\pi $ and $\pi /2$, the system produces solely symmetric modes
[Figs. \ref{SolitonExp}(c) and (d)].

\begin{figure}[tbp]
\centering{\label{fig3a} \includegraphics[scale=0.25]{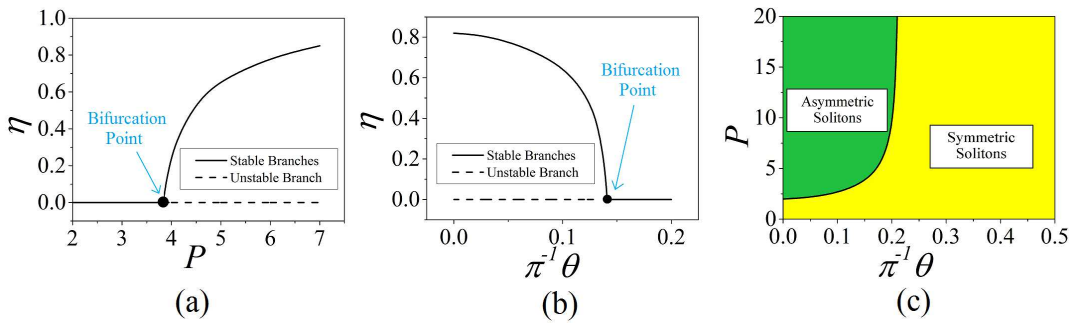}}
\caption{(Color online) (a,b) Symmetry-breaking bifurcation diagrams shown using $\protect\eta (P)$ and $\protect\eta (\protect\theta )$ curves [Fig. (%
\protect\ref{eta})]. In panels (a) and (b), $(J,\protect\theta )=(1,\protect%
\pi /6)$ and $(J,P)=(1,3)$, respectively. Black solid dots are the
bifurcation points; in panel (a), $P=3.845$; in panel (b), $%
\protect\theta =0.141\protect\pi $ (i.e., $\protect\theta \approx 25.4^{\circ
}$). The solutions on the solid curves (stable branches) are stable, whereas those on the dashed lines (unstable branches) are unstable. (c) Black curve is the border between the yellow and green areas,
which are populated by cross-symmetric and asymmetric solitons, respectively, in
the $(P,\protect\theta )$ plane, which is the locus of the bifurcation points for $%
J=1$. The border is composed of supercritical symmetry-breaking bifurcation points. Here, $\protect\sigma =0$ (there are no contact interactions).}
\label{J1}
\end{figure}

%\begin{figure}[tbp]
%\centering{\label{figstable}}\includegraphics[scale=0.2]{4aa.eps}}
%\caption{(Color online) Unstable (symmetric) soliton with $(P,J,\theta)=(5,1,\pi/6)$, which is selected from the unstable branch in Fig. \ref{J1}(a). (b)Direct simulation (real-time evolution) to the soliton solution in panel(a). (c)Stable (asymmetric) soliton with the same controlled parameter in panel (a), i.e. $(P,J,\theta)=(5,1,\pi/6)$, which is selected from the stable branch in Fig.
%\end{figure}
\begin{figure}[tbp]
\centering{\label{figstable} \includegraphics[scale=0.20]{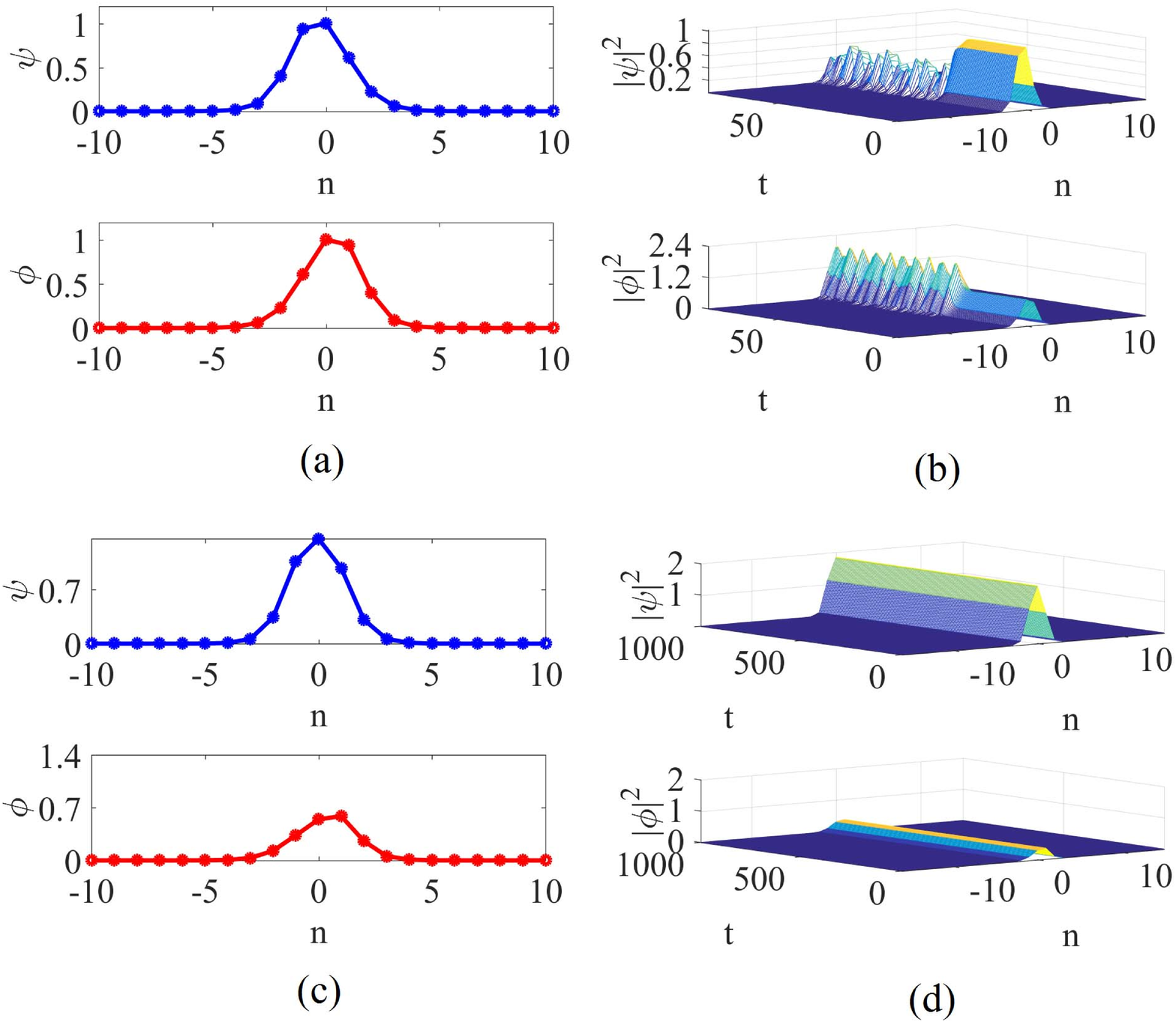}}
\caption{(Color online) Unstable (symmetric) soliton with $(P,J,\theta)=(5,1,\pi/6)$, which is produced by the ITM with a symmetric input. (b) Direct simulation (real-time evolution) of the soliton solution in panel (a). (c) Stable (asymmetric) soliton with the same controlled parameter as in panel (a), i.e., $(P,J,\theta)=(5,1,\pi/6)$, which is produced by the ITM with an asymmetric input. (d) Direct simulation (real-time evolution) of the soliton solution in panel (c).}
\label{stablely1}
\end{figure}

\begin{figure}[tbp]
\centering{\label{fig4a} \includegraphics[scale=0.25]{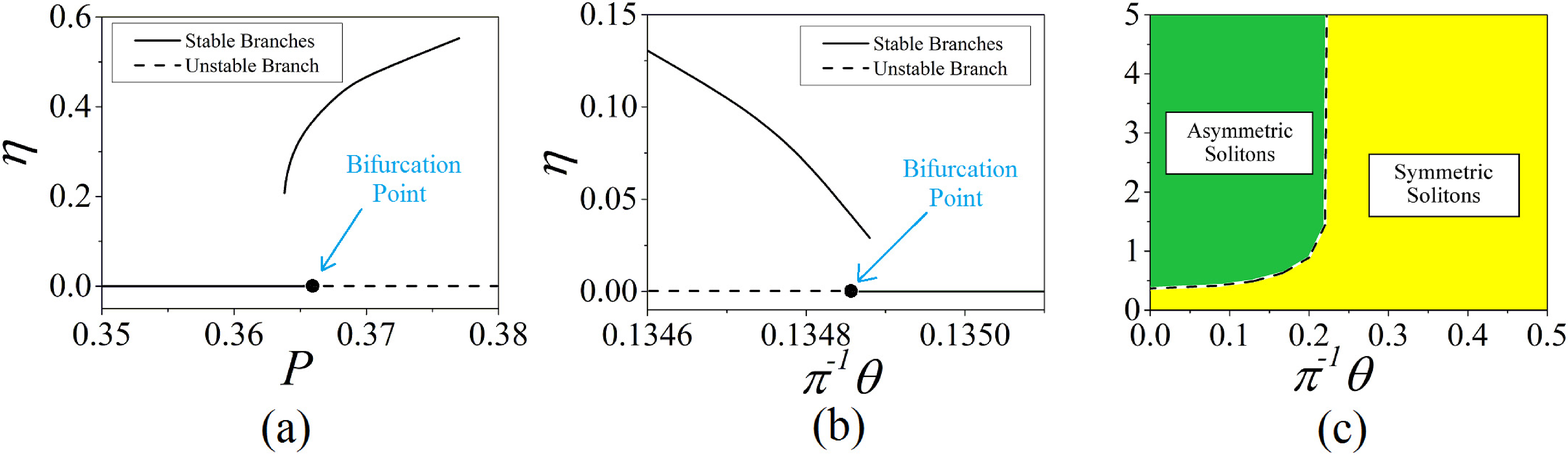}}
\caption{(Color online) The same as Fig. \protect\ref{J1}, but for a much
weaker vertical (interwell) hopping rate, $J=0.1$. In panels (a) and (b), $%
\protect\theta =0$ and $P=0.5$, respectively, are fixed. The bifurcation
points are $P=0.365$ in (a) and $\protect\theta =0.13488\protect\pi $ (i.e.,
$\protect\theta \approx 24.28^{\circ }$) in (b). Backward-going curves,
which must connect the bifurcation point with the stable asymmetric branches
\protect\cite{Sub}, are not shown here, as imaginary-time integration
methods cannot produce these unstable solutions.
(c) Subcritical symmetry-breaking bifurcation (black dashed line). Here and in Fig.
\protect\ref{Trajectories} below, to clearly display the locus of the
bifurcation points, we do not show a bistable area of overlap between
the cross-symmetric and asymmetric solitons, as it is too narrow to be clearly
shown.}
\label{J01}
\end{figure}

\begin{figure}[tbp]
\centering{\label{figstable} \includegraphics[scale=0.20]{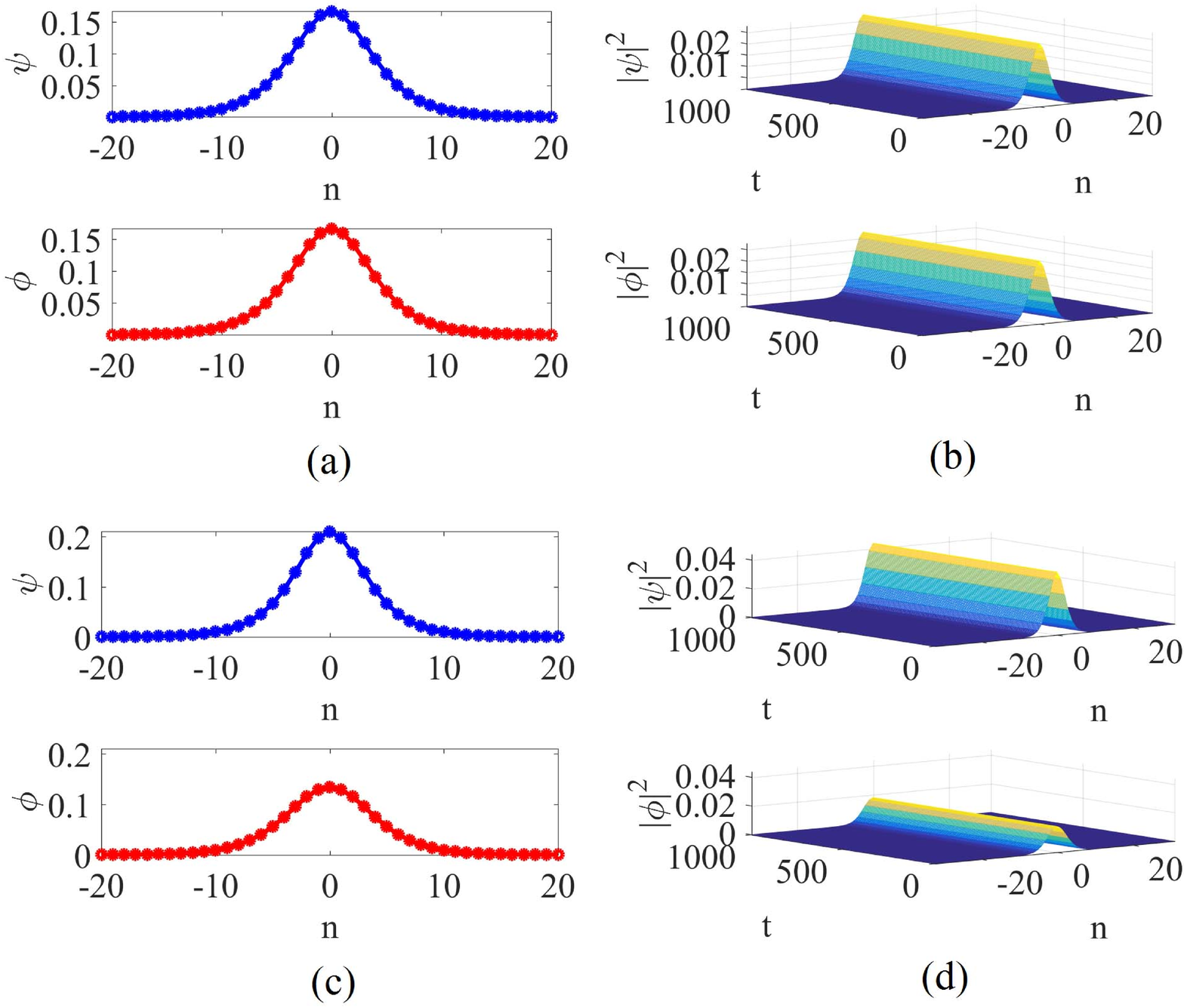}}
\caption{(Color online) Stable (symmetric) soliton with $(P,J,\theta)=(0.365,0.1,0)$, which is produced by the ITM with a symmetric input. (b) Direct simulation (real-time evolution) of the soliton solution in panel (a). (c) Stable (asymmetric) soliton with the same controlled parameter as in panel (a), i.e., $(P,J,\theta)=(0.365,0.1,0)$, which is produced by the ITM with an asymmetric input. (d) Direct simulation (real-time evolution) of the soliton solution in panel (c).}
\label{bistable}
\end{figure}

Figures \ref{J1}(a) and (b) show the bifurcation diagrams of cross-symmetry breaking as plots
of the asymmetry $\eta $, as defined in Eq. (\ref{eta}), versus $P$ at fixed $%
\theta $, and $\eta $ versus $\theta $ at fixed $P$, for asymmetric discrete
solitons at $J=1$, when only the supercritical bifurcation diagram
is found. Typical examples of solutions selected from stable and unstable branches in Fig. \ref{J1}(a) are displayed in Fig. \ref{stablely1}. To summarize these findings, the locus of bifurcation points is plotted in
the $\left( P,\theta \right) $ plane in Fig. \ref{J1}(c). The latter figure
demonstrates that, at $\theta <0.22\pi $ (i.e., $\theta <39.6^{\circ }$),
increasing $P$ leads to supercritical cross-symmetry-breaking bifurcation (i.e., the symmetry-breaking phase transition of the second kind). However, at $\theta >0.22\pi $, cross-symmetry breaking does not occur at arbitrarily large values of norm $P$. Figure \ref{J1}(c) also implies that, at fixed $P>3.845$, one can
induce supercritical cross-symmetry breaking by rotating the dipoles to smaller values of $%
\theta $, i.e., closer to the system's axis. The fact that symmetry breaking never
occurs for sufficiently large $\theta$ can be easily understood
qualitatively. Indeed, with increasing $\theta $, the attractive DDI
in each component (along each sublattice), which drives the symmetry breaking, weakens, whereas the attraction between the sublattices, which obviously tends to suppress the symmetry breaking, strengthens. For this
reason, $\theta =0.22\pi $\ is close to (slightly larger than) angle (%
\ref{theta1}).

Figure \ref{J01} shows the bifurcation diagrams of the symmetry breaking for $J=0.1$, i.e.,
for a much smaller interwell hopping rate. In this case, in contrast to
that considered above for $J=1$, only the subcritical cross-symmetry-breaking bifurcation (i.e., the symmetry-breaking phase transition of the first
kind \cite{Sub}) is found. Figure \ref{bistable} shows typical examples of the bistable soliton in the area where the two stable branches overlap. Similar to the result for $J=1$, at $\theta
>0.22\pi $, cross-symmetry breaking does not occur at arbitrarily large values of $P$.
At $\theta <0.22\pi $ and $P>0.365$, either increasing $P$ or downward
rotation of $\theta $ from $\pi /2$ to $0$ leads to the subcritical cross-symmetry-breaking bifurcation. Notice that the narrow intermediate unstable branches of asymmetric solitons, which connect the stable asymmetric branch and the bifurcation point on the symmetric branch, are missing, as the ITM does not converge to such unstable solutions \cite{YLi2012-2}.

The subcritical bifurcation gives rise to bistability, i.e., a region of
coexistence of stable cross-symmetric and asymmetric discrete solitons \cite%
{Sub}. This region is too narrow to appear clearly in Fig. %
\ref{J01}(c) [a typical bistable case is shown below in Fig. \ref%
{Trajectories}(d)].

\begin{figure}[tbp]
\centering{\label{fig5a} \includegraphics[scale=0.26]{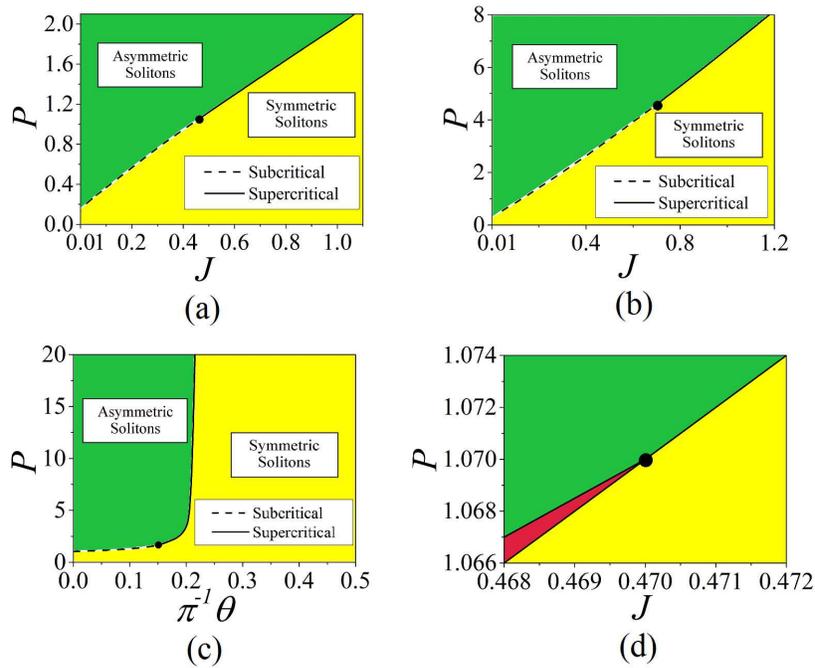}}
\caption{(Color online) (a,b) Loci of bifurcation points in the
$\left( J,P\right) $ plane for fixed values of the dipole orientation
angle $\protect\theta =0$ and $0.196\protect\pi $, respectively. Here and in
other figures, the solid and dashed lines designate
supercritical and subcritical symmetry-breaking bifurcations, respectively. Critical points separating
the super- and subcritical regions in panels (a) and (b) are $%
(P,J)=(1.07,0.47)$ and $(4.59,0.7)$, respectively. (c) Locus of
bifurcation points in the $\left( P,\protect\theta \right) $ plane for a
fixed vertical (interwell) hopping rate, $J=0.45$ [cf. Figs.
\protect\ref{J1}(c) and \protect\ref{J01}(c)]. The critical point separating
the sub- and supercritical regions is $(P,\protect\theta )=(1.63,0.15\protect%
\pi )$. (d) Magnified view of region near the critical point (black dot) in panel
(a). Red wedge is a narrow bistable area where stable
cross-symmetric and asymmetric solitons coexist. The symmetry breaking in this area is of
the subcritical type. }
\label{Trajectories}
\end{figure}

Figures \ref{J1} and \ref{J01} suggest that the type of symmetry-breaking bifurcation,
i.e., the characteristics of the symmetry-breaking phase transition, can be
switched by varying the parameters $(P,J,\theta )$. Figures \ref{Trajectories}%
(a), (b), and (c) demonstrate this possibility by displaying the loci of bifurcation
points in cases where the bifurcation type indeed changes. This is shown in
the $(P,J)$ plane for $\theta =0$ and $0.196\pi $, as well as in the $\left(
P,\theta \right) $ plane for $J=0.45$ (for $\theta
=0$, the cross-symmetric solitons are replaced by the usual symmetric ones,
as stated above). In these figures, solid and dashed lines designate
supercritical and subcritical bifurcations, respectively, which are separated by the critical points
(black dots).

\begin{figure}[tbp]
\centering{\label{fig6a} \includegraphics[scale=0.2]{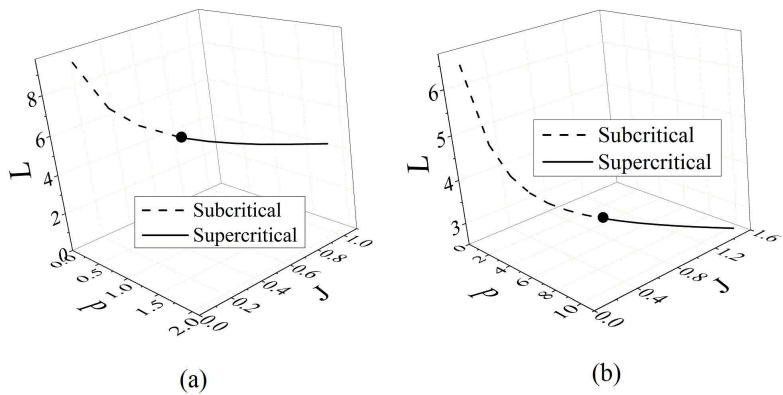}}
\caption{(Color online) Size of the (cross-)symmetric soliton at the
bifurcation point, as defined in Eq. (\protect\ref{L}), versus $P$ and $J$
for $\protect\theta =0$ (a) and $\protect\theta =0.196\protect\pi $ (b). The
critical points between the super- and subcritical segments in panels (a) and
(b) are located at $L=5.431$ and $3.021$, respectively. }
\label{AeffvsJP}
\end{figure}

To characterize the super- and subcritical symmetry-breaking bifurcations in further
detail, we defined the size of the cross-symmetric soliton as
\begin{equation}
L=\frac{\left( \sum_{n}|\psi _{n}|^{2}\right) ^{2}}{\sum_{n}|\psi _{n}|^{4}}
\label{L}
\end{equation}%
(because the cross-symmetric solitons obey the constraint $\psi _{n}=\phi _{-n}$%, as well as $P_{\psi }=P_{\phi }$, it is sufficient to use a single component
to define their size). Figure \ref{AeffvsJP} shows $L$ at the bifurcation
point as a function of $P$ and $J$ for $\theta =0$ and $\theta =\arcsin (1/%
\sqrt{3})$. The subcritical bifurcation tends to occur
in broader solitons (with larger $L$), and the curvature of the $L(P,J)$
line is also larger when the bifurcation is subcritical.

\begin{figure}[tbp]
\centering{\label{fig7a} \includegraphics[scale=0.26]{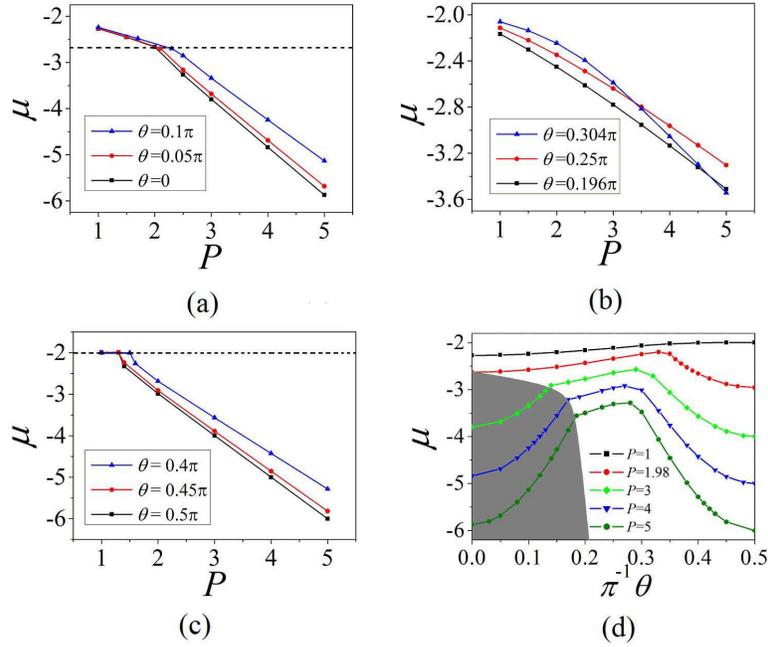}}
\caption{(Color online) (a,b,c) $\protect\mu $ versus $P$ for different fixed values
of $\protect\theta $ in three regions: $\protect\theta <0.196\protect\pi $, $%
0.196\protect\pi <\protect\theta <0.304\protect\pi $, and $\protect\theta %
>0.304\protect\pi $, respectively. In (a), the dashed line designates the
transition to asymmetric solitons via symmetry breaking, whereas in (c), flat segments
correspond to delocalized states in a region where solitons do not exist. In
this figure, $J=1$ and $\protect\sigma =0$. (d) Chemical potential of two-component
discrete solitons, $\protect\mu $, versus $\protect\theta $ at different
fixed values of the total norm $P$. The solitons are asymmetric in the
shaded area. }
\label{mu}
\end{figure}

Families of solitons are usually described by the dependence between
the chemical potential $\mu$ and norm $P$. Figure \ref{mu} shows this dependence,
supplemented by a panel showing $\mu $ as a function of the orientation angle $%
\theta $ at fixed $P$. The $\mu (P)$
dependence behaves somewhat differently in three regions: $%
\theta <0.196\pi $, $0.196\pi <\theta <0.304\pi $, and $\theta >0.304\pi $
[see typical examples for these three cases in Figs. \ref{mu}(a), (b), and (c),
respectively]. Note that in all these panels, the Vakhitov--Kolokolov
criterion, $d\mu /dP<0$, holds; this is well-known as a necessary stability
condition for a bright soliton supported by attractive nonlinearity \cite%
{VK,reviews}. Short flat segments with $\mu \approx -2$ in Fig. \ref{mu}(c)
belong not to solitons but to delocalized states, as the DDI cannot
generate solitons in these regions. For $\theta <0.192\pi $, the
change in the slope of the $\mu (P)$ lines at $\mu \approx -2.7$ [dashed line in Fig. \ref{mu}(a)] corresponds to the transition from
cross-symmetric solitons to asymmetric ones. The smooth shape of $\mu (P)$
in the region $0.192\pi <\theta <0.304\pi $ [Fig. \ref{mu}(b)] is
explained by the absence of symmetry breaking in this case [Fig. \ref{J1}(c)]. The dependence of the $\mu$ value of the two-component soliton on the angle $\theta$ in this system is also studied. Figure \ref{mu}(d) shows the $\mu$ value of the two-component soliton versus $\theta$, namely, $\mu(\theta)$, for different values of $P$. This panel clearly shows that $\mu$ depends sensitively on $\theta$. The relationship between $\mu$ and $P$ shows different behavior in three regions: $\theta<0.196\pi$, $0.196\pi<\theta<0.304\pi$, and $\theta>0.304\pi$.

\subsection{System including contact interactions, $\protect\sigma %
\neq 0$}

\begin{figure}[tbp]
\centering{\label{fig8a} \includegraphics[scale=0.26]{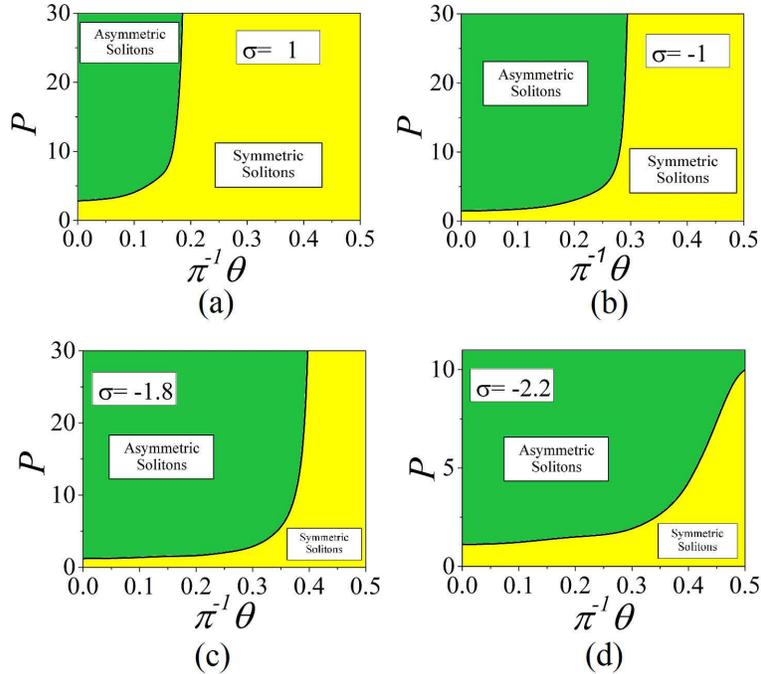}}
\caption{(Color online) Same as Fig. \protect\ref{J1}(c) ($J=1$ in this figure and the following one), but in the presence of the
contact interaction terms in Eq. (\protect\ref{baseEq}) with strength $%
\protect\sigma =1$ (a), $-1$ (b), $-1.8$ (c), and $-2.2$ (d). In all
cases, the symmetry-breaking bifurcation is supercritical. In panels (a), (b),and (c), symmetry breaking does
not occur at $\protect\theta >\protect\theta _{\mathrm{cr}}$, with $%
\protect\theta _{\mathrm{cr}}=0.192\protect\pi $, $0.3\protect\pi $, and $%
0.413\protect\pi $, respectively.}
\label{sigma}
\end{figure}

\begin{figure}[tbp]
\centering{\label{fig9a} \includegraphics[scale=0.26]{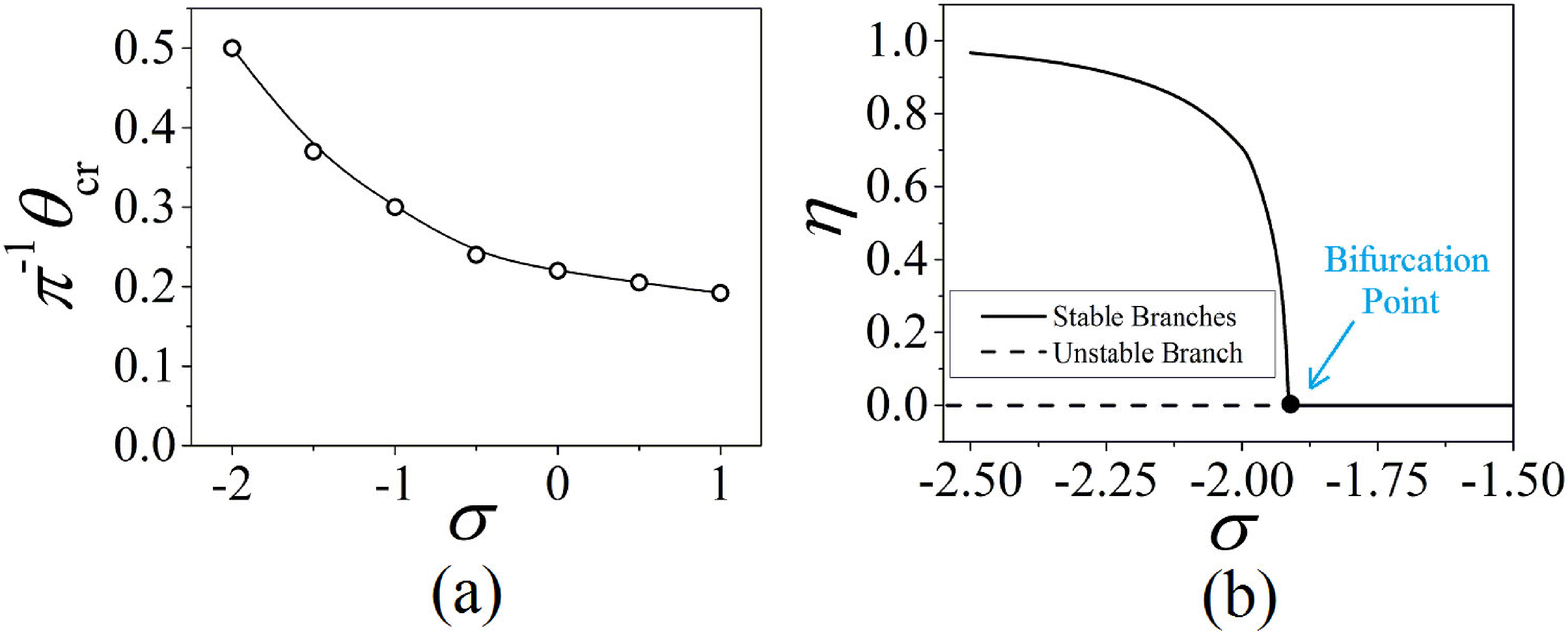}}
\caption{(Color online) (a) $\protect\theta _{\mathrm{cr}}$, which is defined in the
caption of Fig. \protect\ref{sigma}, as a function of the strength $\protect%
\sigma $of the attractive contact interaction. (b) Symmetry-breaking bifurcation diagram in the
form of $\protect\eta $ vs. $\protect\sigma $, where the other parameters are
fixed: $(P,\protect\theta )=(10,0.4\protect\pi )$. The bifurcation point is
located at $\protect\sigma =-1.914$.}
\label{sigma2}
\end{figure}

To check the structural stability of the spontaneous symmetry-breaking phenomena against the
inclusion of additional terms, we here address the contribution from the
local nonlinearity, which is described by the coefficient $\sigma $ in Eq. (\ref%
{baseEq}). This analysis also aims to establish a link between the present
study and previous work that analyzed the spontaneous symmetry breaking in discrete
two-component solitons created by on-site (local) interactions \cite%
{Herring2007}. Figure \ref{sigma} displays areas populated by
cross-symmetric and asymmetric solitons in the $(P,\theta )$ plane for
fixed $J=1$. A comparison with the counterpart of this diagram
at $\sigma =0$ in Fig. \ref{J1}(c) shows that the addition of the moderately
strong repulsive ($\sigma >0$) and attractive ($\sigma <0$) contact
interactions do not dramatically change the picture produced by the DDI. In
particular, the symmetry-breaking bifurcation remains supercritical.

A qualitative change appears in the presence of a sufficiently strong
attractive local nonlinearity, i.e., $\sigma <-2$; as Fig. \ref{J1}%
(d) shows, in this case spontaneous symmetry breaking occurs at all values of $\theta $, whereas at
$\sigma >-2$ (including the case of $\sigma =0$ considered above), symmetry breaking
is absent in the interval $\theta _{\mathrm{cr}}<\theta \leq \pi /2$. In this region, the attractive contact nonlinearity is not enough to overcome the repulsive DDI along each sublattice, which suppresses the emergence of symmetry breaking. The corresponding dependence, $\theta _{\mathrm{cr}}(\sigma )$, which reaches $\pi /2$ and thus actually disappears at $\sigma \approx -2$, is shown in Fig. \ref{sigma2}(a). In addition, Fig. \ref{sigma2}(b) demonstrates that
the onset of the symmetry-breaking bifurcation can be controlled by means of the
contact interaction strength $\sigma $, which, in turn, may be tuned using the Feshbach resonance \cite{FBresonance}.

\section{Conclusion}

We introduced a model of a dipolar BEC consisting of a chain of double-well potential traps. In the tight-binding approximation, it amounts to a system of
two coupled discrete GPEs with long-range DDIs
determined by the angle $\theta $ of the
orientation of the dipoles with respect to the system's axis. Except for the
limiting cases of $\theta =0$ and $\theta =\pi /2$, the system gives rise to
cross-symmetric discrete solitons, the main issue being the phase
transitions (bifurcations) into spontaneous symmetry breaking of
the cross-symmetric solitons. We found that the onset of spontaneous symmetry breaking, as
well as the shape of the asymmetric solitons it generates, is controlled by
the soliton's total norm $P$, rate of vertical (interwell) hopping $J$%, strength $\sigma $ of the contact part of the nonlinearity, and, most
importantly, by $\theta $. In particular, the supercritical symmetry-breaking bifurcation
tends to switch to the subcritical type with decreasing $\theta $ and $J$%.

A challenging issue may be an extension to a 2D discrete two-component
system corresponding to dipolar BECs trapped in a deep 2D lattice
cut into parallel sublattices by a repelling laser sheet. In that geometry,
the DDI will be more complex than that in the 1D system. In particular, symmetry
breaking may be expected not only in fundamental 2D solitons, but also in
discrete solitary vortices \cite{Zhaopin}.

\begin{acknowledgments}
The authors appreciate the very useful discussion with Prof. Boris A. Malomed. This work was supported by the NNSFC (China) through Grant Nos. 11575063, 61471123, and 61575041, and by the Natural Science Foundation of Guangdong Province through Grant No. 2015A030313639.

\end{acknowledgments}

\bibliographystyle{plain}
\bibliography{apssamp}
% Produces the bibliography via BibTeX.

\end{document}